\documentclass[aps,prb,twocolumn,amsmath,amssymb,showpacs,superscriptaddress]{revtex4-1}

\usepackage{graphicx}
\usepackage{amsmath,amssymb,amsfonts}
\usepackage{textcomp}
\usepackage{hyperref}
\usepackage{gensymb}
\usepackage{verbatim}
\usepackage{amsmath}
\hypersetup{breaklinks=true,colorlinks=true,urlcolor=black}
\usepackage{color}
\usepackage{soul}
\DeclareGraphicsExtensions{.jpg,.pdf,.png,.eps}

\begin{document}
\title{Physical properties of $R$Bi$_2$ ($R=$La, Ce) under pressure}
\author{Li Xiang}
\affiliation{Ames Laboratory, Iowa State University, Ames, Iowa 50011, USA}
\affiliation{Department of Physics and Astronomy, Iowa State University, Ames, Iowa 50011, USA}
\email[]{ives@iastate.edu}
\author{Elena Gati}
\affiliation{Ames Laboratory, Iowa State University, Ames, Iowa 50011, USA}
\affiliation{Department of Physics and Astronomy, Iowa State University, Ames, Iowa 50011, USA}
\author{Kathryn Neilson}
\affiliation{Department of Physics and Astronomy, Iowa State University, Ames, Iowa 50011, USA}
\author{Sergey L. Bud'ko}
\affiliation{Ames Laboratory, Iowa State University, Ames, Iowa 50011, USA}
\affiliation{Department of Physics and Astronomy, Iowa State University, Ames, Iowa 50011, USA}
\author{Paul C. Canfield}
\affiliation{Ames Laboratory, Iowa State University, Ames, Iowa 50011, USA}
\affiliation{Department of Physics and Astronomy, Iowa State University, Ames, Iowa 50011, USA}

\email[]{canfield@ameslab.gov}

\date{\today}

\begin{abstract}
We present a study of electrical transport properties of $R$Bi$_2$ ($R$ = La, Ce) under hydrostatic pressure up to $\sim$ 2.5 GPa. These measurements are complemented by thermodynamic measurements of the specific heat on CeBi$_2$ at different pressures up to 2.55 GPa. For CeBi$_2$, we find a moderate increase of the antiferromagnetic transition, $T_\text N$, from 3.3 K to 4.4 K by pressures up to 2.55 GPa. Notably, resistance measurements for both CeBi$_2$ and LaBi$_2$ show signatures of superconductivity for pressures above $\sim$ 1.7 GPa. However, the absence of superconducting feature in specific heat measurements for CeBi$_2$ indicates that superconductivity in CeBi$_2$ (and most likely LaBi$_2$ as well) is not bulk and likely originates from traces of Bi flux, either on the surface of the plate-like samples, or trapped inside the sample as laminar inclusions.
\end{abstract}

\maketitle 

\section{Introduction}
%Bi compounds have been a great platform to study spin-orbit coupling due to Bi-6$p$ electrons and its role in establishing a variety of interesting ground states\cite{Anna2013}. For example, Bi-based families such as $A$Bi ($A$ = Li and Na)\cite{Sambongi1971,Kushwaha2014}, $A$Bi$_2$ ($A$ = K, Rb, Cs and Ca)\cite{Roberts1976,Winiarski2016} and $A$Bi$_3$ ($A$ = Sr, Ba, Ca, Ni, Co, and La) are superconducting (SC) at low temperature \cite{Matthias1952PR,Shao2016,Kinjo2016,Xiang2018,Gati2018,Tence2014}. $R$Bi ($R$ = Ce, Nd, Tb and Dy) and $R$Bi$_2$ ($R$ = La-Nd, Sm) families establish magnetic ground states\cite{Nereson1971,Petrovic2002}. Moreover, Bi-rich compounds have recently received lots of attention as candidates for realizing novel topological phases such as Bi$_2$Se$_3$ and Bi$_2$Te$_3$\cite{Hasan2010,Qi2011,Hor2010,Xia2009,Chen2009}.

Bi-rich compounds manifest a rich variety of ground states. For example, Bi-based families such as $A$Bi ($A$ = Li and Na)\cite{Sambongi1971,Kushwaha2014}, $A$Bi$_2$ ($A$ = K, Rb, Cs and Ca)\cite{Roberts1976,Winiarski2016} and $A$Bi$_3$ ($A$ = Sr, Ba, Ca, Ni, Co, and La) are superconducting (SC) at low temperature \cite{Matthias1952PR,Shao2016,Kinjo2016,Xiang2018,Gati2018,Tence2014}. $R$Bi ($R$ = Ce, Nd, Tb and Dy) and $R$Bi$_2$ ($R$ = La-Nd, Sm) families have low-temperature magnetic ground states with complex $H - T$ phase diagrams\cite{Nereson1971,Petrovic2002}. Moreover, due to the strong spin-orbit coupling of Bi-6$p$ electrons they can have substantial ferromagnetic anisotropy, like MnBi\cite{Anna2013,Taufour2015}, or, more recently, they have became candidates for realizing novel topological phases, such as topological insulators or topological superconductors\cite{Hasan2010,Qi2011,Hor2010,Xia2009,Chen2009}.

Among these, the $R$Bi$_2$ family displays different magnetic ground states depending on the choice of $R$\cite{Petrovic2002}. Structurally, $R$Bi$_2$ forms in an orthorhombic structure with single layers of Bi separated from each other by $R$Bi bilayers that are stacked along the crystallographic $b$ axis\cite{Petrovic2002,Zhou2018PRB}. When $R$ is chosen to be the moment-bearing Ce ion, an antiferromagnetic (AFM) ground state below $T_\text N \sim$ 3.3 K can be stabilized\cite{Petrovic2002}. A recent study shows that CeBi$_2$ is a Kondo system with a Sommerfeld coefficient $\gamma$ over 200 mJ/mol K$^2$ and Kondo temperature of an order of $\sim $ 2 K\cite{Zhou2018PRB}. On the other hand, for $R=$ La (non-moment bearing), LaBi$_2$ reveals metallic behavior without indications of magnetic ordering or superconductivity down to 1.8 K\cite{Petrovic2002}.

In this study, we perform a comparative study of the ground-state tunability of these two members by external pressure. We explore the temperature-pressure phase diagram of CeBi$_2$ and LaBi$_2$ by resistance measurements and complement these, in case of CeBi$_2$, with specific heat measurements. Our results show that $T_\text N$ of CeBi$_2$ is moderately increased upon increasing pressure. Surprisingly, resistance measurements of both CeBi$_2$ and LaBi$_2$ show signatures pressure-induced superconductivity at low temperature ($T\lesssim$ 4 K) above very similar threshold pressures ($p\gtrsim$ 1.68 GPa). However, specific heat measurement of CeBi$_2$ does not reveal any anomaly that could be associated with a transition into the superconducting state. We assign these effects to filamentary SC that likely originates from traces of Bi flux, either on the surface of the plate-like samples, or trapped inside the sample as laminar inclusions. Finally, the analysis of pressure-dependent resistance data at fixed temperatures for CeBi$_2$ suggests that there might be a pressure-induced crossover most likely associated with pressure-induced changes in the Kondo temperature and crystal electric field splitting.

%In this work, we present a detailed pressure study on CeBi$_2$ up to 2.55 GPa by utilizing both resistance and ac specific heat measurements. Our results reveal that the antiferromagnetic transition $T_\text N$ increases from 3.3 K to 4.4 K up to 2.55 GPa. Above 1.68 GPa, signatures of superconductivity is observed below $\sim$ 5 K in resistance measurements. However, our ac specific heat measurements under pressure imply that the observed superconductivity is non-bulk. As comparison, resistance of LaBi$_2$ under pressure up to 2.52 GPa is also studied. Signatures of superconductivity in resistance measurements at the similar pressure range is observed as well. Furthermore, an analysis of pressure dependent resistance for both CeBi$_2$ and LaBi$_2$ suggests possible pressure-induced phase transitions at $\sim$ 1.9 GPa and $\sim$ 1.4 GPa, respectively.

%Bi-rich compounds have recently received lots of attention as candidates to search for novel topological phases due to the strong spin-orbit coupling which is associated with Bi-6$p$ electrons\cite{Hasan2010,Qi2011,Hor2010,Anna2013}. 

\section{Experimental details}
Single crystals of CeBi$_2$ and LaBi$_2$ were grown by a Bi self-flux technique with the help of a frit-disk alumina Canfield Crucible Set\cite{Canfield1992,Canfield2016a}.
%The resulting crystals are mm size(Fig. \ref{XRD}) and were characterized by x-ray using a Rigaku MiniFlex diffractometer (Cu $K_{\alpha \text{1,2}}$ radiation) at room temperature on ground single crystals.
For CeBi$_2$, Ce and Bi in the molar ratio 9:91 were loaded into a crucible set and sealed into a fused silica ampoule under partial argon atmosphere. The ampoule was heated to 1000 $^\circ$C in 5 h and dwelled at this temperature for another 4 h. It was then slowly cooled to 600 $^\circ$C over 45 h. At this temperature, the ampoule was removed from the furnace and excess liquid was decanted by the help of a centrifuge. For LaBi$_2$, La and Bi in the molar ratio 8:92 were loaded into the crucible set, heated to 1000 $^\circ$C in 5h, dwelled at 1000 $^\circ$C for 2 h, and slowly cooled to 350 $^\circ$C over 80 h. The resulting crystals of CeBi$_2$ and LaBi$_2$ are millimeter-size and plate-shaped. Both CeBi$_2$ and LaBi$_2$ crystals are air-sensitive, the preparation of experiments was therefore performed in a N$_2$ glovebox.

The $ac$, in-plane resistance measurements were performed in a Quantum Design Physical Property Measurement System (PPMS) using a 1 mA excitation with frequency of 17 Hz, on cooling using a rate of - 0.25 K/min. The magnetic field was applied perpendicular to the current direction. For CeBi$_2$, two different samples (labeled as S1 and S2) were used in resistance measurements. S1 was measured at ambient condition outside pressure cell and S2 was measured under pressure. The temperature-dependent resistance data for S1 is normalized by extrapolating $p\leq 1.23 GPa$ pressure-dependent resistance data, $R(p)$, at 300 K from S2 back to 0 GPa (see Fig. \ref{fig1_RT}). For LaBi$_2$, only one sample was measured under pressure with the pressures 0.60 GPa $\leq p \leq$ 2.52 GPa. For both compounds, a standard four-contact configuration was used with contacts made by Dupont 4929N silver paint. Specific heat measurements under pressure were performed using an ac calorimetry technique on a third sample (sample S3) in a cryogen-free cryostat from ICEOxford (Lemon-Dry) with base temperature of 1.4 K. Details of the setup used and the measurement protocol are described in Ref. \onlinecite{Gati2019}.

In this study, a Be-Cu/Ni-Cr-Al hybrid piston-cylinder cell, similar to the one described in Ref. \onlinecite{Budko1984}, was used to apply pressure. Good hydrostatic conditions were achieved by using a 4:6 mixture of light mineral oil:n-pentane as pressure medium, which solidifies, at room temperature, in the range $3-4$ GPa, i.e., well above our maximum pressure\cite{Budko1984,Kim2011,Torikachvili2015}. Pressure values were inferred from the $T_{c}(p)$ of lead\cite{Bireckoven1988}, determined via resistance measurements.

%\begin{figure}
%	\includegraphics[width=8.6cm]{XRD}%
%	\caption{Powder x-ray diffraction pattern (black line) and reported peak positions of Bi (red), (a) CeBi$_2$ and (b) LaBi$_2$. Insets: pictures of CeBi$_2$ and LaBi$_2$ crystals.
%		\label{XRD}}
%\end{figure}

\section{Results}
\subsection{CeBi$_2$}
\begin{figure}
	\includegraphics[width=8.6cm]{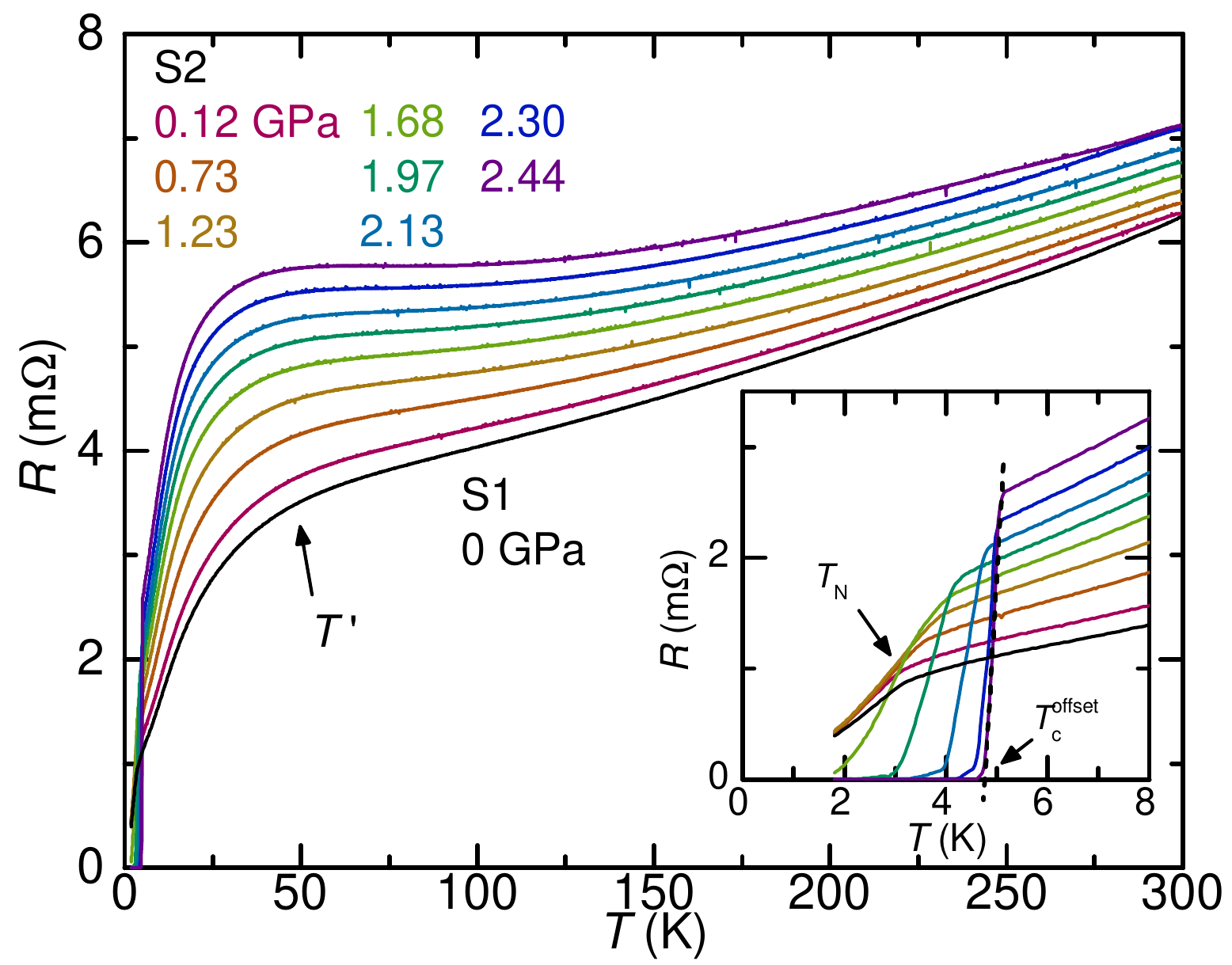}%
	\caption{Resistance of CeBi$_2$ as a function of temperature at ambient pressure (measured on sample S1) and at different finite pressures up to 2.44 GPa (measured on sample S2). The ambient pressure data for S1 is normalized by extrapolating $p\leq$ 1.23 GPa pressure-dependent resistance data, $R(p)$, at 300 K from S2 back to 0 GPa. A broad hump feature is present in all data sets. The inferred crossover temperature $T '$ is exemplarily marked for the data set at 2.44 GPa (for more details, see text). Inset: Blowup of the resistance data at low temperatures showing the magnetic and superconducting transitions. AFM transition temperature $T_\text N$ is indicated by arrow. Criterion for $T_\textrm{c}^\textrm{offset}$ is indicated by arrow.
		\label{fig1_RT}}
\end{figure}

Figure \ref{fig1_RT} shows the temperature-dependent resistance of CeBi$_2$ at ambient pressure (sample S1) and pressure up to 2.44 GPa (sample S2).
%As described in the experimental details, S1 was measured at ambient pressure outside the pressure cell, whereas S2 was directly mounted onto the pressure cell feedthrough and measured under pressure. 
The temperature-dependent resistance data for S1 is normalized by extrapolating the 300 K pressure-dependent resistance data ($R(p)$ for $p\leq$ 1.23 GPa) measured from S2 back to 0 GPa. As shown in the figure, the resistance decreases upon cooling, showing a metallic behavior. At $T\sim$ 50 K, a broad drop of resistance is observed. In an earlier work, it was suggested that this drop in $R(T)$ is associated with either the coherence in Kondo scattering or crystal electric-field (CEF) splitting of Ce atoms\cite{Zhou2018PRB}. At $T\sim$ 3.3 K, the resistance shows a kink-like anomaly due to loss of spin-disorder scattering as CeBi$_2$ undergoes an AFM transition at $T_\text N$\cite{Petrovic2002,Zhou2018PRB}. Sample S2 was measured under pressure and at lowest pressure (0.12 GPa), resistance of S2 shows very similar feature as S1. Upon increasing pressure, the resistance gradually increases over a large temperature range (essentially everywhere in the paramagnetic state). This is a typical behavior for a Ce-based Kondo lattice under pressure\cite{Thompson1994,Hegger2000,Nicklas2001,Nicklas2003}. The broad drop of resistance at ambient pressure becomes progressively more pronounced, as pressure is increased, and evolves into a local maximum at highest pressures. The temperature of this broad drop/hump feature is labeled as $T '$ and indicated by arrow in the figure (see below for the description of the criterion used). The evolution of this feature will be analyzed and discussed in more details below. As we move to the low-temperature region (inset to Fig. \ref{fig1_RT}), for $p\le$ 1.23 GPa, the kink-like anomaly, which is associated with the magnetic transition\cite{Petrovic2002,Zhou2018PRB}, is shifted to higher temperatures upon increasing pressure. Even with this slight increase in $T_\text N$, the loss of spin disorder scattering below $T_\text N$ remains fundamentally the same. As a result, the resistance at 1.8 K, $R$(1.8 K), does not show a significant change. Upon increasing from 1.23 GPa to 1.68 GPa, $R$(1.8 K) shows a sudden decrease. For $p>$ 1.68 GPa, the resistance as a function of temperature, $R(T)$, undergoes a much sharper drop and reaches a zero value, suggesting a pressure-induced superconducting phase at low temperature. The critical temperature of this phase is increased upon increasing pressure.

\begin{figure}
	\includegraphics[width=8.6cm]{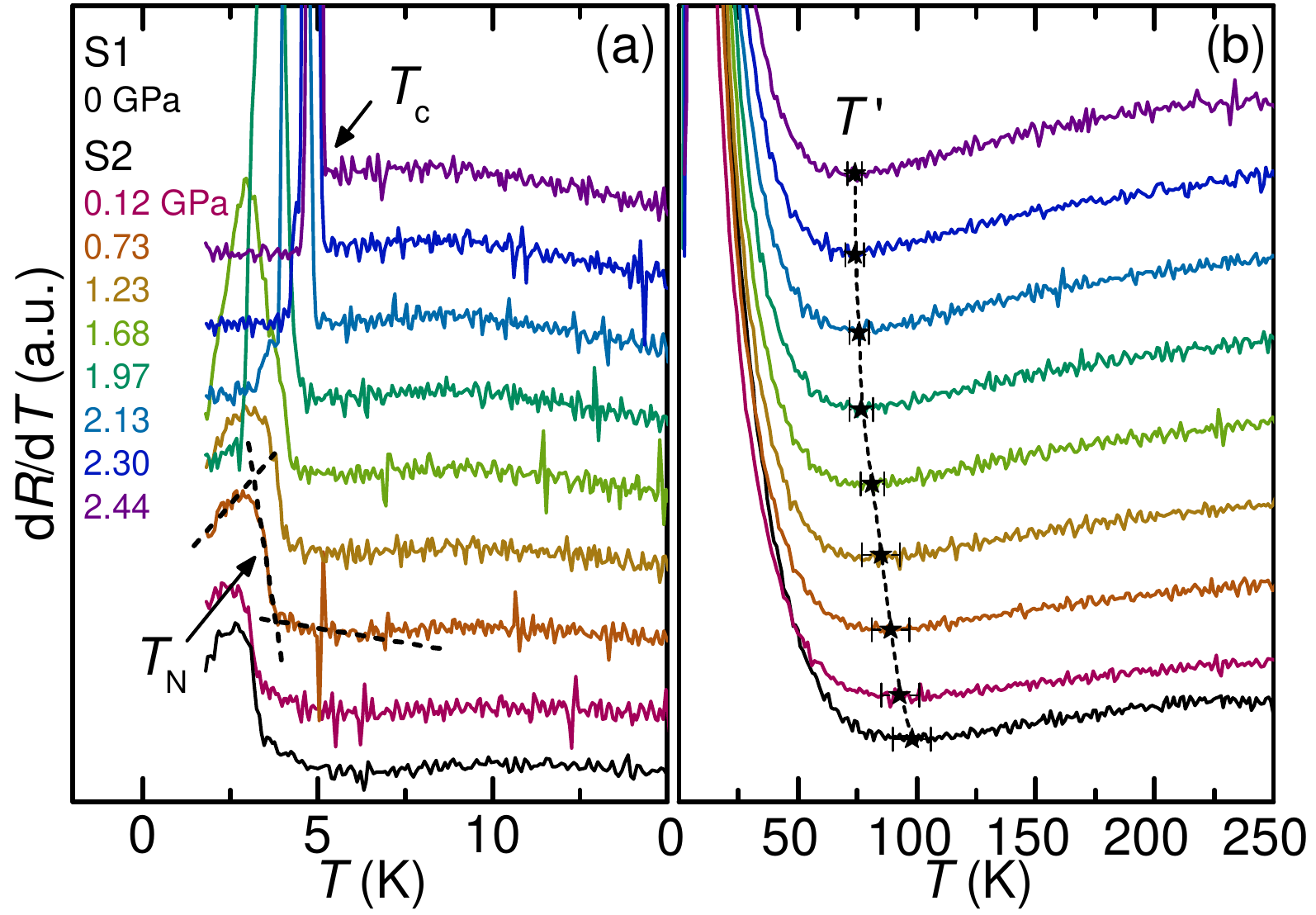}%
	\caption{(a) Temperature derivative of the resistance,d$R$/d$T$, in the low-temperature region ($T\le$ 15 K). The criterion for the determination of the AFM transition temperature $T_\text N$ is illustrated by dashed lines and marked by the arrow (midpoint of the jump-like feature). At high pressures, the magnetic anomaly is masked by a strong drop of resistance, likely due to spurious SC (see main text). The respective temperature is denoted by $T_\text c$ (see arrow). (b) Temperature derivative of the resistance,d$R$/d$T$, showing the evolution of the temperature associated with the broad hump feature in $R(T)$ curves. $T'$ is determined by the minimum in d$R$/d$T$ curves. Data sets in (a) and (b) are offset for clarity.
		\label{dRdT_fig3}}
\end{figure}

The temperature-derivative of the resistance data is shown in Fig, \ref{dRdT_fig3} to better differentiate between the low $p$ and high $p$ feature at low temperature as well as to trace the broad feature at $T \sim$ 50 K. As shown in Fig.\ref{dRdT_fig3} (a), at low pressures ($p\le$ 1.23 GPa), the magnetic transition shows up as a jump-like feature in the d$R$/d$T$. We therefore define $T_\text N$ as the midpoint of the jump-like feature in d$R$/d$T$ (see dotted lines and arrow in Fig. \ref{dRdT_fig3} (a) as well as Figs. \ref{fig3_cT} (b) and (c) below). As a result, $T_\text N$ increases with increasing $p$ with a slope of $\sim$ 0.48 K/GPa. At higher pressures ($p\ge$ 1.68 GPa), the superconducting transition can be seen as a sharp peak in d$R$/d$T$. Figure \ref{dRdT_fig3} (b) shows d$R$/d$T$ curves over a larger temperature range. As shown in the figure, the broad drop/hump features in $R(T)$ are reflected in minima d$R$/d$T$. We therefore define the crossover temperature $T '$, which marks the change between two different resistance regimes, by the minima in the the d$R$/d$T$ as indicated by the dashed lines in the figure. It is clearly seen that $T '$ increases upon increasing pressure.

\begin{figure}
	\includegraphics[width=8.6cm]{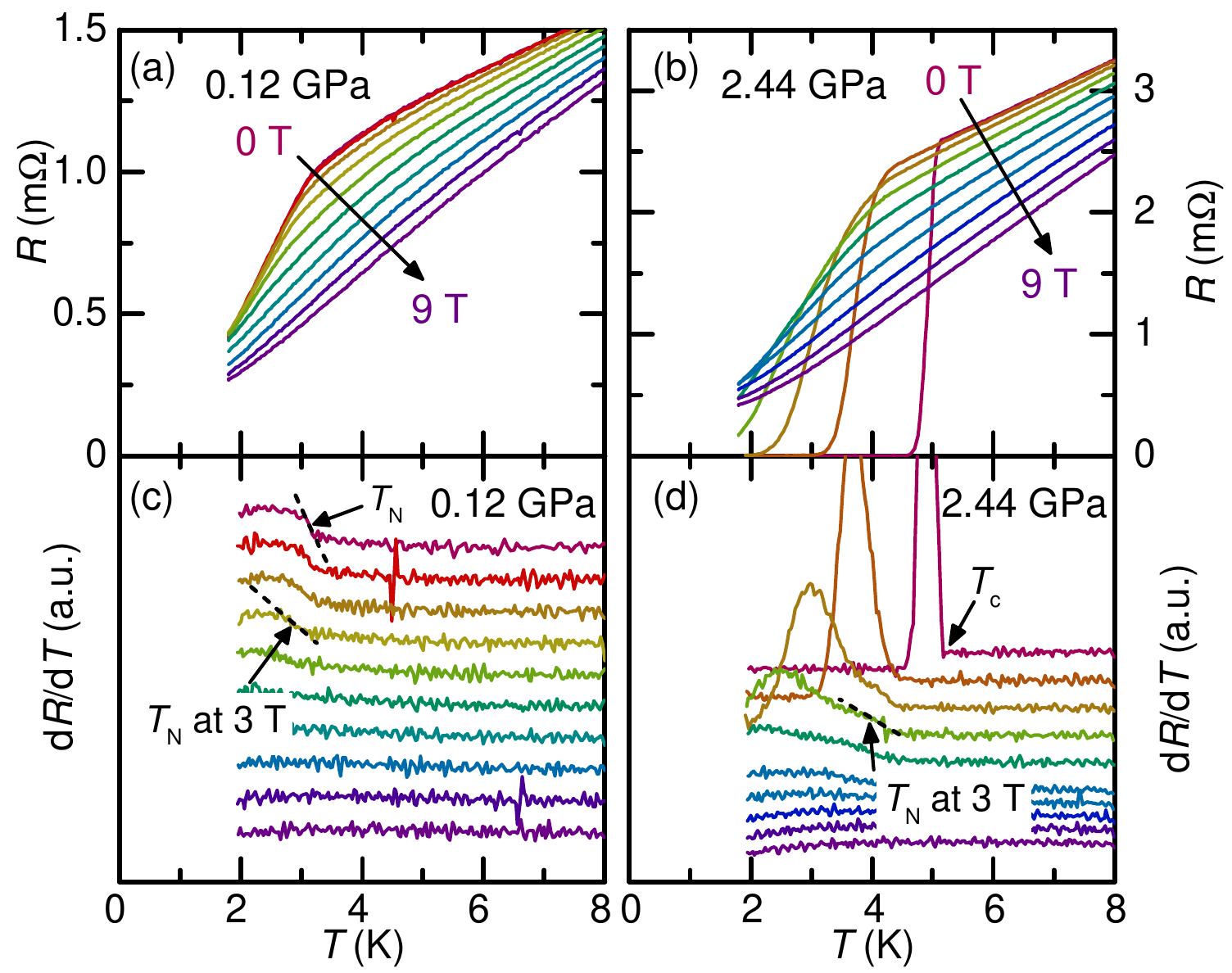}%
	\caption{(a), (b) Temperature-dependent resistance of CeBi$_2$ S2 under magnetic fields up to 9 T for selected pressures. Fields are applied along the $b$-axis. (c), (d) Temperature-derivative of the resistance data, taken in applied magnetic fields, shown in (a) and (b), respectively. Data sets are offset for clarity. Criteria for $T_\text N$ at 0 T and 3 T are indicated by arrows (midpoint of the jump-like feature).
		\label{fig2_RTH}}
\end{figure}

To trace the magnetic transition to higher pressures, the temperature-dependent resistance under magnetic fields up to 9 T applied along the $b$-axis was studied. The applied field can suppress the superconducting transition which masks the signature of the magnetic transition for $p\ge$ 1.68 GPa. The results for selected pressures are presented in Fig. \ref{fig2_RTH}. As shown in Figs. \ref{fig2_RTH} (a) and (c), at 0.12 GPa the kink-like anomaly in $R(T)$ associated with magnetic transition is broadened in higher fields, yet not much shifted with an applied field of 3 T. In the temperature derivative of the resistance data, the corresponding jump-like feature is suppressed with increasing magnetic fields until it disappears at higher fields. At 2.44 GPa, the sharp drop of the resistance in $R(T)$ associated with superconducting transition at $\sim$ 5 K is suppressed to lower temperatures with magnetic fields and the kink-like anomaly re-emerges at $\sim$ 4 K. Further increasing magnetic fields broadens the kink-like anomaly until it disappears. Similarly, in the temperature derivative d$R$/d$T$, we first observed a sharp peak associated with the superconducting transition at low magnetic fields. Upon increasing the field, the sharp peak is suppressed and shifted to lower temperatures, at the same time, a second jump-like feature emerges. At even higher fields, both features disappear. By analogy we associate this re-emerged kink-like anomaly in $R(T)$ (jump-like feature in d$R$/d$T$) with the same magnetic transition that is observed at low pressures. The resistance does not become zero at 1.8 K for magnetic field $B\geq$ 2 T indicating a critical field of $\sim$ 2 T at 1.8 K.

\begin{figure}
	\includegraphics[width=8.6cm]{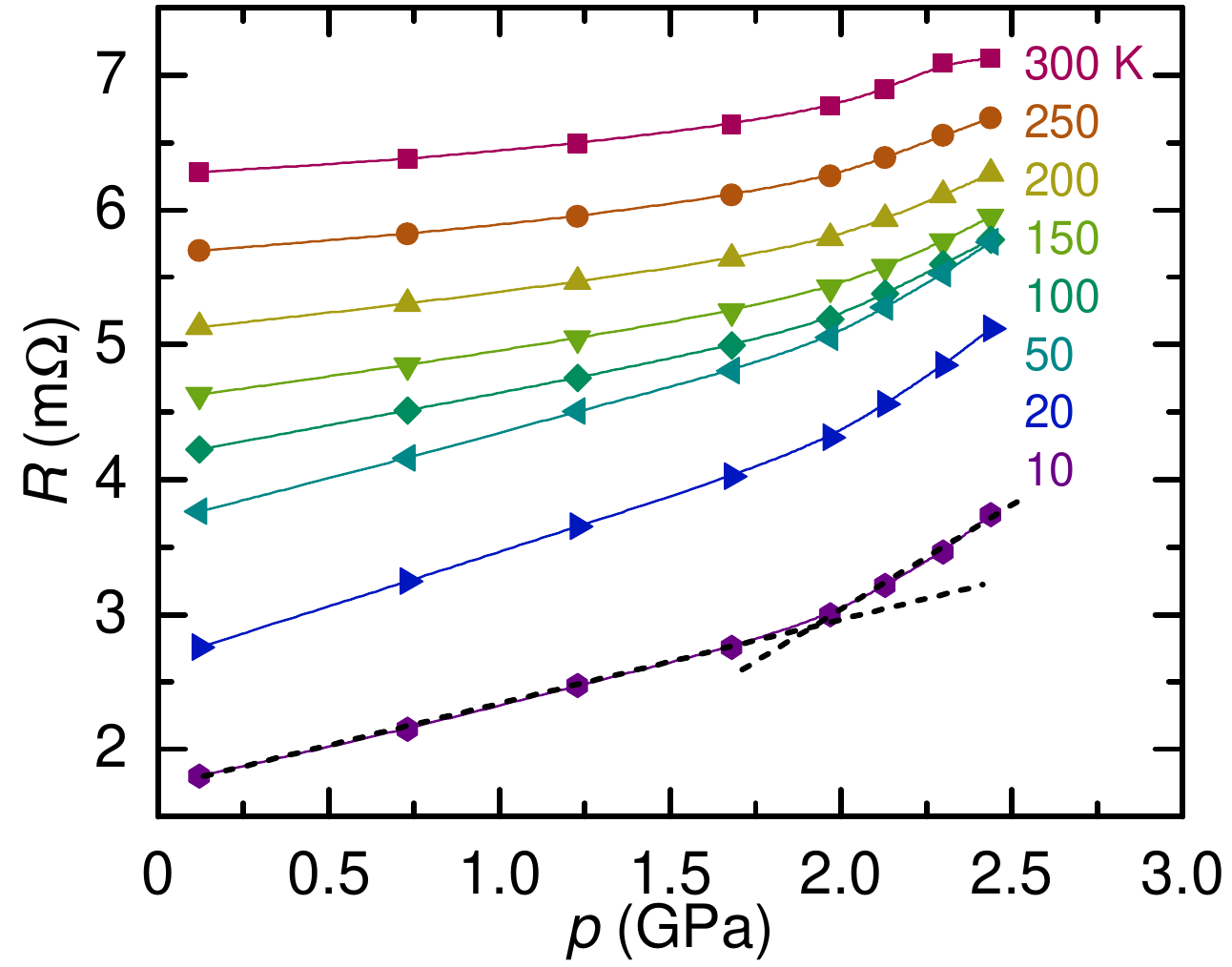}%
	\caption{Pressure dependence of resistance, $R(p)$, at fixed temperatures for CeBi$_2$. A change of slope between 1.68 GPa and 1.97 GPa is indicated by the cross of the dashed line.
		\label{RP_CeBi2}}
\end{figure}

To further investigate the overall increase of resistance with pressure, we present in Fig. \ref{RP_CeBi2} the pressure dependent resistance $R(p)$ at fixed temperatures. As shown in the figure, a change of slope is observed when pressure is increased from 1.68 GPa to 1.97 GPa at 10 K, this feature persists up to 300 K, the highest temperature investigated in this study. The strongest pressure responses are for $T\lesssim T '$, suggesting shifts in the Kondo feature around $T '$. Whereas the $R(p)$ data for 300 K are quite similar to what is found for LaBi$_2$ in Fig. \ref{RP_LaBi2} (see below).

\begin{figure}
	\includegraphics[width=8.6cm]{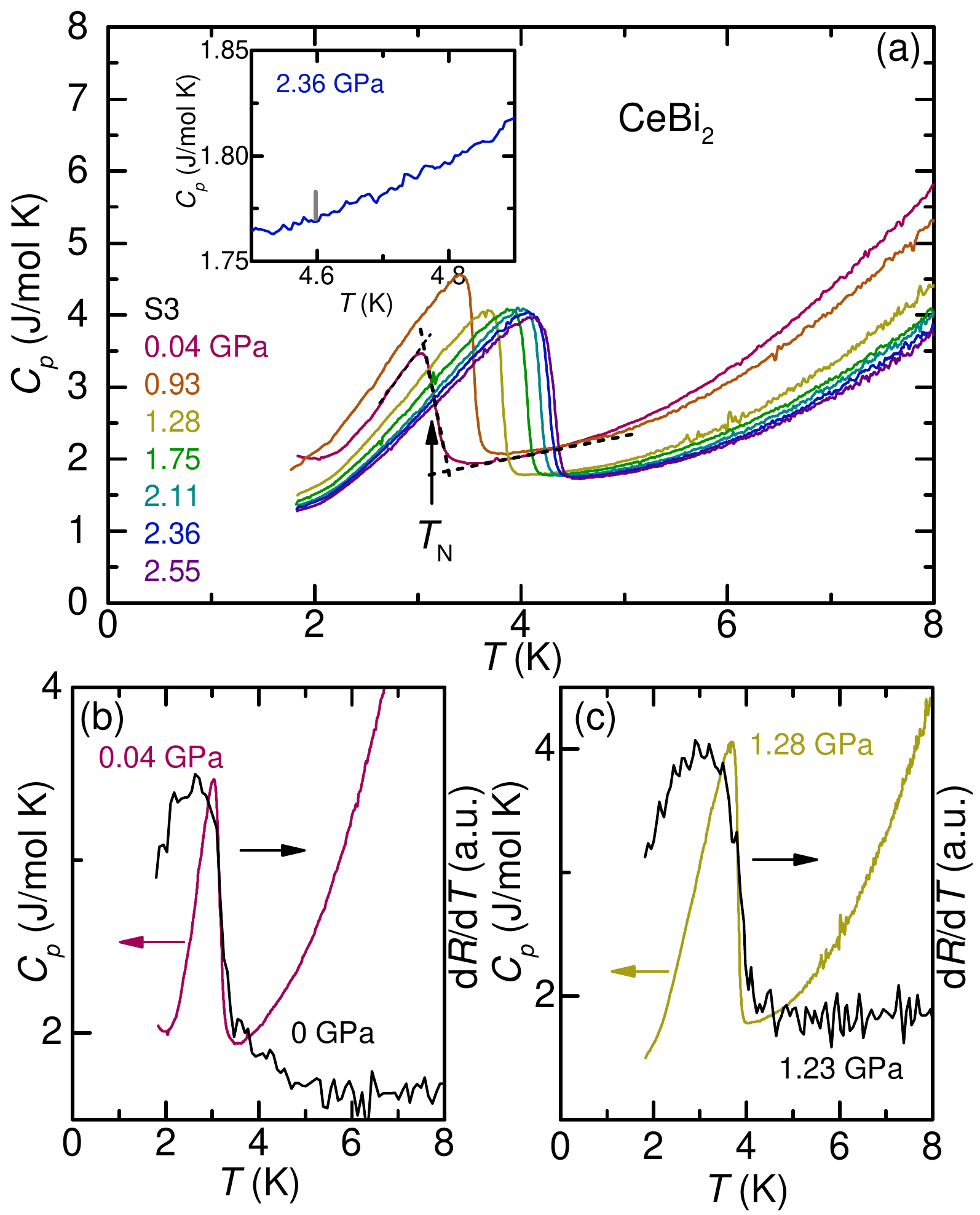}%
	\caption{(a) Evolution of the temperature-dependent specific heat, $C_p(T)$, with pressure up to 2.55 GPa for CeBi$_2$ S3. Criterion for $T_\text N$ is indicated by arrow (midpoint of the specific heat jump). The inset shows the data near 4.6 K for 2.36 GPa, the gray vertical line indicates a 13.2 mJ/mol K specific heat jump at 4.6 K (details are discussed in the main text). (b), (c) Temperature-dependent specific heat data and temperature-derivative of the resistance data at two sets of nearly identical pressures ((b) 0.04 GPa and 0 GPa, (c) 1.28 GPa and 1.23 GPa). Note that the midpoint criterion gives same $T_\text N$ values for both data sets.
		\label{fig3_cT}}
\end{figure}

The observation of a state with zero resistance in CeBi$_2$ calls for a thermodynamic investigation of the temperature-pressure phase diagram. Thus, we studied the specific heat of CeBi$_2$ (sample S3) under pressure and the results are presented in Fig. \ref{fig3_cT} (a). At lowest pressure (0.04 GPa), very close to ambient pressure, the specific heat, $C_p(T)$, nicely reveals a nearly mean-field-like anomaly at $T\sim$ 3.2 K, which speaks in favor of a second-order phase transition. The shape, position, and size of the feature is consistent with the specific results of a previous study and therefore allows us to assign this feature to the magnetic transition at $T_\text N$. Figures. \ref{fig3_cT} (b) and (c) show the comparison between temperature dependent $C_p$ and d$R$/d$T$ at two sets of nearly identical pressures (0.04 GPa and 0 GPa, 1.28 GPa and 1.23 GPa). As shown in the figure, temperature-dependent $C_p(T)$ and d$R$/d$T$ exhibit similar jump-like feature at the transition temperature which is consistent with the Fisher-Langer relation\cite{Fisher1968,Alexander1976}. Thus, to determine the transition temperature, $T_\text N$, from specific heat measurement, same criterion as in the resistance measurement is used (midpoint of jump-like anomaly as indicated by dashed lines and arrow in Fig. \ref{fig3_cT} (a)).
As pressure is increased up to 2.55 GPa, $T_\text N$ is monotonically increased. At the same time, the jump size of the anomaly does not significantly change indicating that the amount of entropy released at $T_\text N$ is unchanged. However, we did not observe a second feature at any pressure, thus suggesting that CeBi$_2$ does not undergo any other phase transition than the magnetic one. This includes in particular also a possible superconducting transition for $p>$ 1.68 GPa inferred from our resistance data. One might argue that a possible superconducting feature in specific heat is masked by the huge entropy release at the magnetic transition, as $T_\text N$ and the resistive $T_\text c$ are very close. However, even at high pressure, at which we expect that $T_\text N$ and $T_\text c$ are well separated, no feature in specific heat occurs (see inset of Fig. \ref{fig3_cT} (a)). Another possibility for the apparent absence of a specific heat feature might be that the superconducting jump size is very small and therefore falls below the resolution limit. In the following, we provide estimates for the lower and upper bound of superconducting jump size in CeBi$_2$.

For a phonon-mediated BSC superconductor, the specific heat jump at the superconducting transition can be written as,
\begin{equation}
	\Delta C = 1.43\gamma T_\text c,
	\label{C jump}
\end{equation}
where $\gamma$ is the electronic Sommerfeld coefficient and $T_\text c$ is the superconducting transition temperature. To estimate a possible lower limit of $\Delta C$, we first assume that superconductivity is unrelated to the Kondo-lattice-nature of CeBi$_2$. Thus, for the choice of $\gamma$, we refer to the nonmagnetic reference LaBi$_2$ which is isostructural to CeBi$_2$. Since LaBi$_2$ has a $\gamma$ value of 2 mJ/mol K$^2$\cite{Petrovic2002}, with $T_\text c$ $\sim$ 4.6 K from Fig. \ref{fig1_RT}, we get $\Delta C = $13.2 mJ/mol K. Compared to the noise level, such value of specific jump (gray vertical line in the inset of Fig. \ref{fig3_cT} (a)) should be resolvable. For an upper limit, we take the $\gamma$ value of the Kondo-lattice CeBi$_2$, 200 mJ/mol K$^2$\cite{Zhou2018PRB}, we get $\Delta C = $1.32 J/mol K, which would be one hundred times larger than the gray vertical line in the inset of Fig. \ref{fig3_cT} (a). The absence of any resolvable specific heat jump feature, which can be associated with superconductivity, suggests that the pressure-induced superconductivity is likely filamentary rather than bulk. This conclusion will be related to again below after presentation of data on LaBi$_2$.

\begin{figure}
	\includegraphics[width=8.6cm]{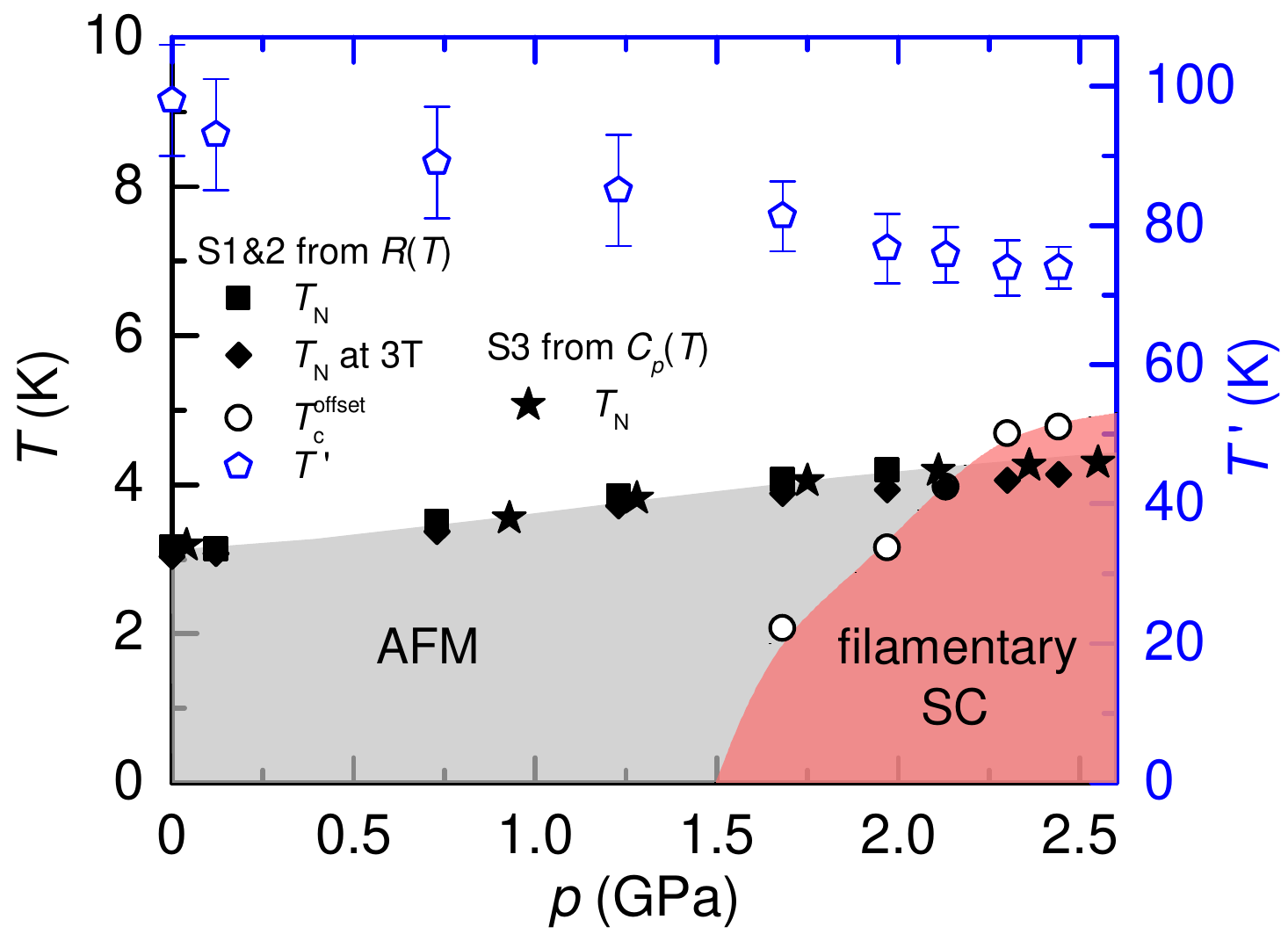}%
	\caption{Temperature-pressure phase diagram of CeBi$_2$ as determined from resistance and specific heat measurements. Black squares and diamonds represent the magnetic transition $T_\text N$ determined from resistance measurement for 0 T and 3 T respectively. Black stars represent $T_\text N$ determined from specific heat measurement. Black open symbols represent the superconducting transition $T_\textrm{c}^\textrm{offset}$ determined from resistance measurement. Blue pentagons represent $T '$ determined from resistance measurement (Note the right axis used here for $T '$). Gray and red areas represent the antiferromagnetically ordered and filamentary-superconducting regions, respectively.
		\label{fig4_phasediagram}}
\end{figure}

We summarize our $T_\text N$ and $T '$ data for CeBi$_2$ as well as our $T_\text c^\text {offset}$ (filamentary) data in the temperature-pressure ($T-p$) phase diagram shown in Fig. \ref{fig4_phasediagram}. For the magnetic transition, both $T_\text N$ at zero field and 3 T from resistance measurement (Fig. \ref{fig2_RTH}) and $T_\text N$ from zero field specific heat data are included. For superconducting transition, $T_\text c^\text {offset}$ is determined from resistance measurement (Fig. \ref{fig1_RT} (b)). The $T_\text N$ values, inferred from $R(T,p)$ and $C(T,p)$ agree reasonably well within their experimental resolution. As shown in Fig. \ref{fig4_phasediagram}, magnetic field suppresses magnetic transition $T_\text N$ slightly ($\sim$ 0.2 K by 3 T), as is often the case for antiferromagnets. $T_\text N$ increases monotonically with pressure with a rate of 0.48 K/GPa up to 2.55 GPa. For superconductivity, it first sets in at $\sim$ 1.68 GPa with a sharp drop in $R(T)$, yet not give rise to zero resistance down to 1.8 K. Upon increasing pressure, the drop in $R(T)$ becomes progressively sharper and zero resistance at low temperature is reached as well. Furthermore, from 1.68 GPa to 2.44 GPa, $T_\text c^\text {offset}$ monotonically increases from 2.1 K to 4.8 K, appearing to saturate at our highest pressure. Finally, the temperature $T '$ associated with Kondo coherence scattering or CEF splitting is suppressed upon increasing pressure, with $T '\simeq$ 98 K at 0 GPa and 74 K at 2.44 GPa.

\subsection{LaBi$_2$}

\begin{figure}
	\includegraphics[width=8.6cm]{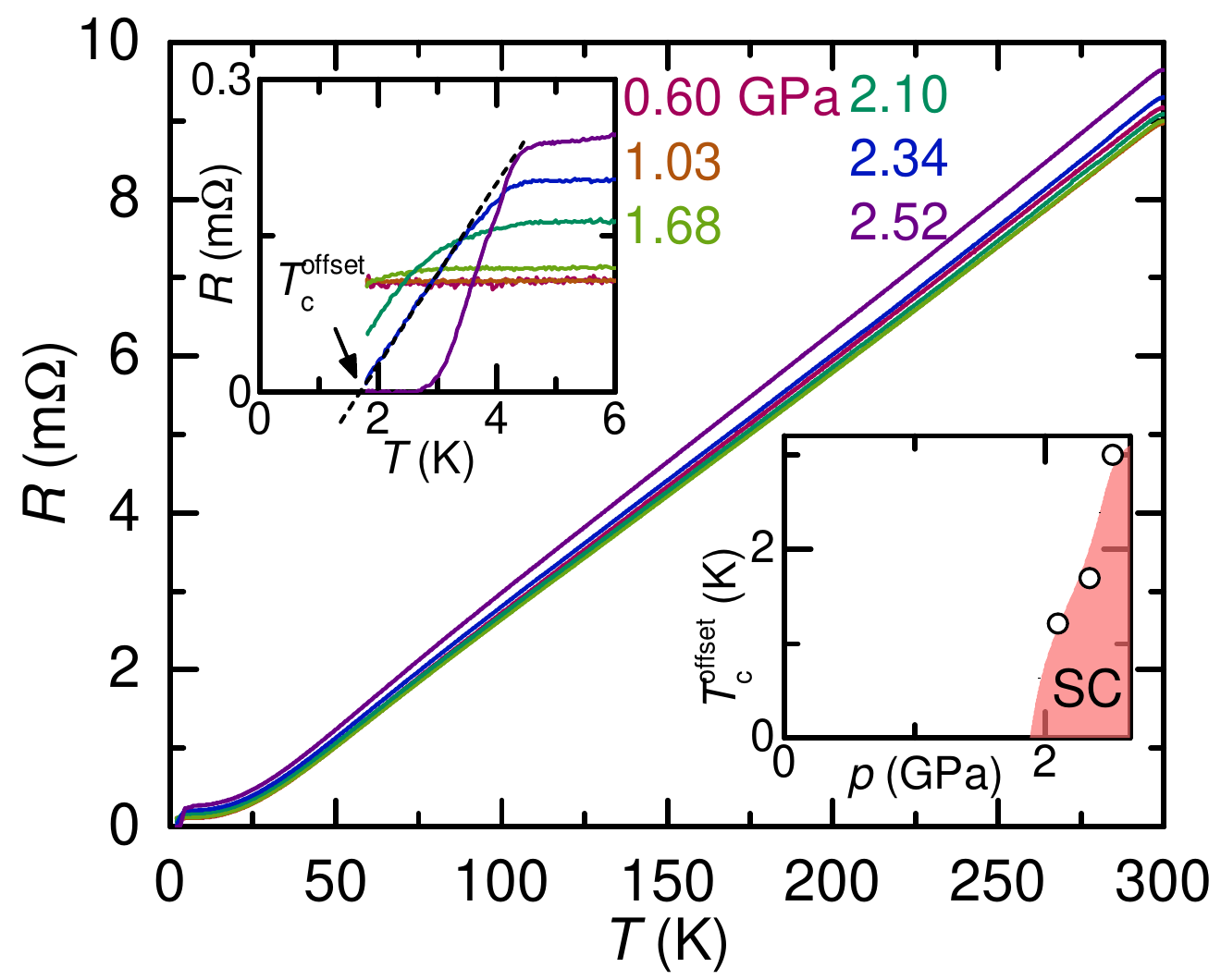}%
	\caption{Resistance of LaBi$_2$ as a function of temperature at different pressures for 0.60 GPa $\leq p \leq$ 2.52 GPa. Upper inset: blowup of the resistance data at low temperatures showing the superconducting transition. Criterion for $T_\text c^\text {offset}$ is indicated by arrow. Bottom inset: superconducting transition temperature, $T_\text c^\text {offset}$, as a function of pressure. Red area represent the superconducting region as inferred from resistance measurement.
		\label{fig5_RT}}
\end{figure}

Next, we discuss our resistance data for the non-magnetic, LaBi$_2$, member of the $R$Bi$_2$ family. Figure \ref{fig5_RT} presents the pressure evolution of the temperature-dependent resistance for LaBi$_2$ with pressures 0.60 GPa $\leq p \leq$ 2.52 GPa.  For all pressures, resistance decreases upon cooling, showing metallic behavior. For a large temperature range ($T \gtrsim$ 50 K), the resistance shows linear dependence on temperature. In the low-temperature region (upper inset of Fig. \ref{fig5_RT}), for $p \le$ 1.03 GPa, resistance as a function of temperature is relatively flat suggesting that the low-temperature resistance is dominated by impurity scattering. At 1.68 GPa, $R(T)$ shows a faster drop of resistance below $\sim$ 2.5 K. When pressure is further increased, this drop of resistance becomes more pronounced. At 2.52 GPa, resistance actually drops to zero below 2.7 K, suggesting pressure-induced superconductivity. The drop of resistance, visible for 1.68 GPa $\le p \le$ 2.34 GPa, is likely to be associated with traces of superconducting phase. Using the criterion defined in the upper inset of Fig. \ref{fig5_RT}, the superconducting transition temperature, $T_\text c^\text {offset}$, can be traced and the results are shown in the bottom inset of Fig. \ref{fig5_RT}. As shown in the figure, $T_\text c^\text {offset}$ increases from 1.2 K to 3 K when pressure is increased from 2.10 GPa to 2.52 GPa.
%\textcolor{red}{Debye temperature 160 K\cite{Petrovic2002}}

\begin{figure}
	\includegraphics[width=8.6cm]{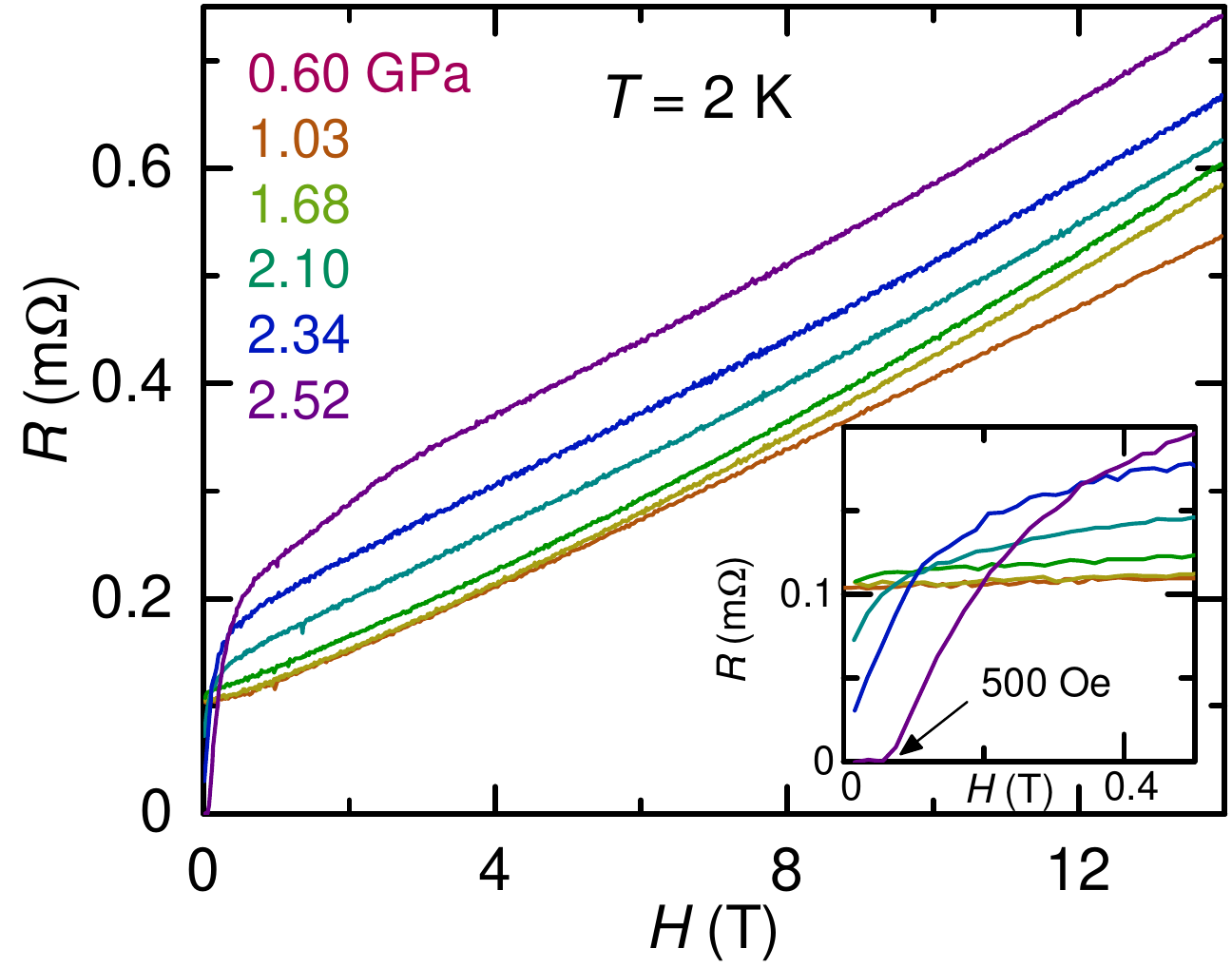}%
	\caption{Evolution of the field-dependent resistance at 2 K of LaBi$_2$ with pressure 0.60 GPa $\leq p \leq$ 2.52 GPa and fields applied along the $b$-axis. The lowest pressure data (0 GPa) is not included due to excessive noise. Inset shows the blowups of the low-field region.
		\label{fig6_RH}}
\end{figure}

The field dependence of the resistance at 2 K was studied and is presented in Fig. \ref{fig6_RH}. For $p\le$1.03 GPa, resistance gradually increases with magnetic field with a slightly up-bending curvature. For $p\ge$1.68 GPa, at low fields, the resistance first undergoes a fast increase upon increasing fields, which is likely due to the suppression of superconductivity. At higher fields, $R(H)$ curves behave similarly with the ones at lower pressures. Moreover, at 2.52 GPa the zero resistance at 2 K is lifted for $H\gtrsim$ 500 Oe, indicating a critical field of $\sim$ 500 Oe. Bearing in mind that close to ambient pressure the magnetoresistance clearly deviates from the conventional $H^2$ behavior, we observe that pressures up to $\sim$ 2.5 GPa do not modify this behavior (besides the lower field effects of superconductivity) in any conspicuous way. The data in Figs. \ref{fig5_RT} and \ref{fig6_RH} are consistent with traces of SC phase, with distributions of $T_\text c$ values existing in the LaBi$_2$ sample. The mean $T_\text c$ of these filamentary traces increases with pressure for $p>$ 1.68 GPa.

\begin{figure}
	\includegraphics[width=8.6cm]{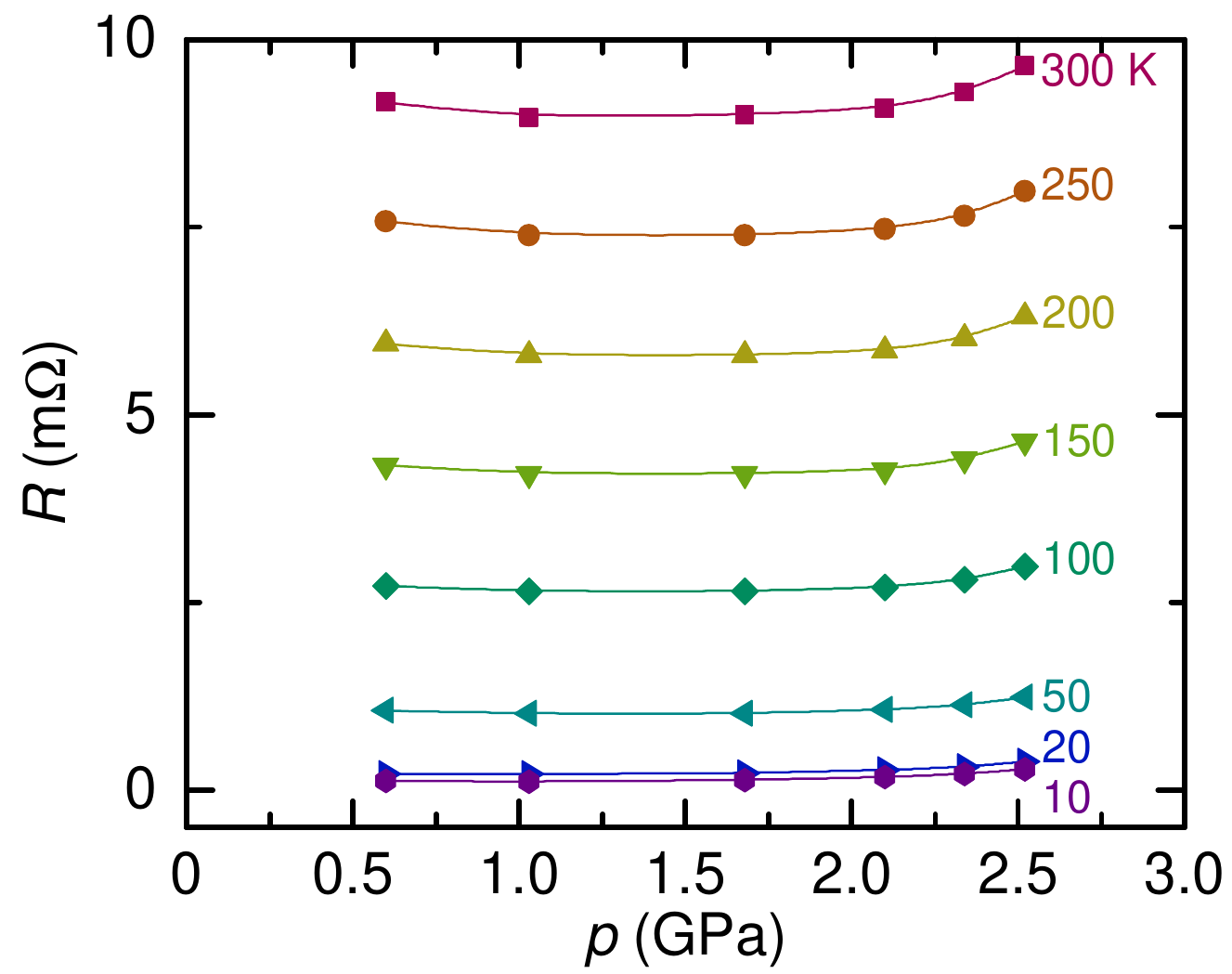}%
	\caption{Pressure dependence of resistance, $R(p)$, at fixed temperatures for LaBi$_2$. The lowest pressure data (0 GPa) is not included due to excessive noise.
		\label{RP_LaBi2}}
\end{figure}

To better visualize the pressure evolution of the higher temperature resistance for LaBi$_2$, Fig. \ref*{RP_LaBi2} presents the pressure-dependent resistance $R(p)$ at fixed temperatures. The resistance of LaBi$_2$ first decreases and then increases with pressure, giving rise to a broad minimum between 1.03 GPa and 1.68 GPa. Compared with the $R(p)$ of CeBi$_2$, $R(p)$ of LaBi$_2$ has a similar higher-pressure, higher-temperature up-turn, but lacks the larger $T\lesssim T '$ pressure dependence seen in CeBi$_2$.

\section{Discussion}
Before discussing the implications of the zero-resistive state, which we observed in CeBi$_2$ and LaBi$_2$ at higher pressures, we first focus on the increase of $T_\text N$ and decrease of $T '$ under pressure in CeBi$_2$, as this is robustly established by our resistance and specific heat study. The properties of a Kondo lattice system are usually dominated by two characteristic energy scales, which are both susceptible to externally applied pressure: Ruderman-Kittel-Kasuya-Yosida (RKKY) interaction energy $T_\text {RKKY} \propto J^2$ and Kondo interaction energy $T_\text K \propto \text e^{-1/J}$ where $J$ is the exchange interaction\cite{Ruderman1954PhyRev,Kasuya1956,Yosida1957PhyRev,Kondo1964,Hewson1993}. When $T_\text {RKKY} \gg T_\text K$, the ground state is magnetic and for $T_\text K \gg T_\text {RKKY}$, it is nonmagnetic. The competition between them and the resulting ground state is often described by the Doniach phase diagram\cite{Doniach1977}. For Ce-based compounds, the ground state is often magnetic. Applying external pressure can suppress magnetic transition temperature to zero and lead to non-magnetic ground state via a quantum critical point\cite{Steglich1979,Jaccard1992,Mathur1998,Park2006,Knebel2006,Jiao2015}. In our study, the AFM transition temperature $T_\text N$ of CeBi$_2$ is moderately increased by pressure up to $\sim$ 2.5 GPa. This suggests that at ambient pressure, CeBi$_2$ is deeply in its magnetic state and higher pressure is needed to suppress $T_\text N$\cite{Knebel2006a,Chen2006,Kimura2007,Bauer2010}. This is compatible with the Doniach picture, as there is a maximum of $T_\text N$ due to the explicit functional dependences of $T_\text {RKKY}$ and $T_\text K$. Moreover, in the Doniach picture, when pressurizing a Ce-based Kondo lattice, an increase of $T_\text K$ is often observed due to the enhancement of exchange interaction $J$\cite{Thompson1994,Goltsev2005}. This, in turn, should give rise to a shift of broad resistive features, associated with $T_\text K$, to higher temperatures with pressure. Therefore, a suppression of $T '$ observed in this study suggests that the broad drop/hump feature in $R(T)$ can not be explained by only the Kondo coherence scattering\cite{Hegger2000,Muramatsu2001,Nicklas2003}.

The resistance measurements for both CeBi$_2$ and LaBi$_2$ reveal a zero-resistive state at high pressures, suggesting a pressure-induced SC phase for these compounds. By comparing their $T-p$ phase diagrams (Figs. \ref{fig4_phasediagram} and \ref{fig5_RT} (a) inset), we see that the two phase diagrams exhibit similar SC phase regions, but with slightly different onset pressures and $T_\text c$ values. For CeBi$_2$ $T_\text c$ saturated at $\sim$ 4.8 K by 2.44 GPa whereas $T_\text c$ of LaBi$_2$ reaches $\sim$ 3 K but seems still rising with pressure. Moreover, at the highest pressures in this study (2.44 GPa for CeBi$_2$ and 2.52 GPa for LaBi$_2$), CeBi$_2$ and LaBi$_2$ have very different critical fields at $\sim$ 2 K ($\sim$ 2 T for CeBi$_2$ and $\sim$ 500 Oe for LaBi$_2$).

Despite the zero-resistive state and relative sharp resistance drop at high pressures for CeBi$_2$ and LaBi$_2$, we would like to argue that the observed SC feature is extrinsic for the following reasons. First of all, specific heat measurement under pressure for CeBi$_2$ does not reveal any SC feature which strongly speaks in favor of filamentary SC. Second, similar $T_\text c$ values for Ce and La are unlikely in bulk $R$Bi$_2$. On one hand, if the SC in these two compounds is standard BSC SC, then hybridizing rare earths such as Ce or Yb suppresses $T_\text c$ aggressively\cite{Maple1972,Canfield1998,Budko2006}. On the other hand, if CeBi$_2$ at high pressures becomes a heavy fermion superconductor, the specific heat jump anomaly at $T_\text c$ should be even bigger. Then similar SC onset pressure and $T_\text c$ between LaBi$_2$ and CeBi$_2$ are unlikely again as LaBi$_2$ is not a heavy fermion compound.
%As for the observed SC under pressure, our resistance measurement for both CeBi$_2$ and LaBi$_2$ show signatures of pressure-induced SC. SC sets in at $p\sim$ 1.68 GPa by having a faster decrease of $R(T)$ in the low-temperature region or a sudden drop of $R$(1.8 K). Resistance further drops to zero value at low temperature at 2.44 GPa and 2.52 GPa for CeBi$_2$ and LaBi$_2$, respectively. Applying magnetic fields lifts the zero resistance and suggesting a critical field of $\sim$2 T and $\sim$500 Oe at 2 K for CeBi$_2$ (at 2.44 GPa) and LaBi$_2$ (at 2.52 GPa), respectively. However, ac specific heat measurement of CeBi$_2$ under pressure up to 2.55 GPa does not reveal any feature that can be associated with this SC state. In other words, up to 2.55 GPa, the SC for CeBi$_2$ observed in resistance measurements is non-bulk. 

To speculate about the possible origin of the filamentary SC, we refer to literature. First we notice that similar situation has been found in other Bi compounds as well where SC is attributed to Bi flux or thin films of Bi\cite{Thamizhavel2003,Mizoguchi2011,Lin2013}. Moreover, it is know that single-crystalline Bi undergoes sequential structural transitions upon increasing pressure and possesses rich physics under pressure\cite{Klement1963,Degtyareva2004,Li2017a}. Specifically, at low temperature, Bi-II exists between 2.55 GPa and 2.70 GPa with $T_\text c \sim$ 3.9 K and upper critical field $\mu_0 H_{c2}$(2 K)$\sim$ 0.05 T, Bi-III exists between 2.70 GPa and 7.7 GPa with $T_\text c \sim$ 7 K and $\mu_0 H_{c2}$(2 K)$\sim$ 3 T\cite{Li2017a}. Owing to the very similar $T_\text c$ of Bi-II to our results on CeBi$_2$ in the almost identical pressure range, we suspect that the filamentary SC we observed in the resistance measurement of CeBi$_2$ originates from traces of Bi flux. It is likely that the SC in LaBi$_2$ is non-bulk and origins from Bi flux as well. Slight differences in onset pressure and  $\mu_0 H_{c2}$ could arise from details of the unit cell parameters which could give rise to slightly different strain conditions.

\section{Conclusion}
In conclusion, the resistance of $R$Bi$_2$ ($R$ = La and Ce) under pressure up to $\sim$ 2.5 GPa and ac specific heat of CeBi$_2$ under pressure up to 2.55 GPa have been studied. Our studies show that for CeBi$_2$ the antiferromagnetic transition temperature, $T_\text N$, increases upon increasing pressure with the rate of $\sim$ 0.48 K/GPa. This fits into the Doniach phase diagram and suggests that there might be a maximum of $T_\text N$, followed by its decrease and finally a quantum critical point at possibly significantly higher pressures. Resistance and ac specific heat measurements of CeBi$_2$ together suggest that the pressure-induced superconductivity in CeBi$_2$ is likely not bulk. It is likely that the SC phase is filamentary Bi either on the surface or as laminar in the bulk of the sample. We suspect the pressure-induced superconductivity in LaBi$_2$ to arise from a similar extrinsic origin giving that the onset pressure and transition temperature of superconductivity are very similar to that of CeBi$_2$. 
Further pressure-dependent resistance analyses for CeBi$_2$ and LaBi$_2$ indicate some anomalies in the $R(p)$ curves, a change of slope between 1.68 GPa and 1.97 GPa for CeBi$_2$ and a broad minimum between 1.03 GPa and 1.68 GPa for LaBi$_2$. Taken together, these suggest that the stronger, low-temperature features see near and below $T '$ for CeBi$_2$ are related to the pressure dependent hybridization and crystal electric field splitting of the Ce.

Finally, we would like to point out, again, that when studying the properties of Bi-rich compounds under pressure, one needs to be very careful and mindful for the various phases elemental Bi has and the rich physics they display at different pressures\cite{Klement1963,Degtyareva2004,Li2017a}.

\begin{acknowledgements}
Work at the Ames Laboratory was supported by the U.S. Department of Energy, Office of Science, Basic Energy Sciences, Materials Sciences and Engineering Division. The Ames Laboratory is operated for the U.S. Department of Energy by Iowa State University under Contract No. DEAC0207CH11358. L.X. was supported, in part, by the W. M. Keck Foundation and the Gordon and Betty Moore Foundation’s EPiQS Initiative through Grant GBMF4411.
\end{acknowledgements}

\clearpage

\bibliographystyle{apsrev4-1}
%\bibliography{MyRef}
%

\end{document}